\documentstyle[eqsecnum,floats,preprint,aps]{revtex}
\tighten

\def\Journal#1#2#3#4{{#1} {\bf #2}, #3 (#4)}

\def\AP{\em Ann. Phys.}
\def\AAS{\em Ann. Acad. Sci.}

\def\NPB{{\em Nucl. Phys.} B}
\def\PLB{{\em Phys. Lett.}  B}

\def\PRD{{\em Phys. Rev.} D}

\def\NCA{\em Nuovo Cimento A}
\def\be{\begin{equation}}
\def\ee{\end{equation}}
\def\bea{\begin{eqnarray}}
\def\eea{\end{eqnarray}}

      \def\Ov{{\mit\Omega}_{\rm v}}

      \def\GeV{\,{\rm GeV}}

           \def\mx{m_{\rm x}}        \def\mc{m_{\rm c}}
 \def\mP{m_{\rm P}\!}       \def\mp{m_{\rm p}}          
 \def\Ev{E_{\rm v}}         \def\lv{\ell_{\rm v}}     \def\Rv{R_{\rm v}}
          \def\nv{n_{\rm v}}        
 \def\nua{\nu_\star}        
 \def\rhv{\rho_{\rm v}}     \def\rhb{\rho_{\rm b}}    
 \def\rhc{\rho_{\rm c}}     \def\rhN{\rho_{_{\rm N}}}
 \def\aa{{g^*}}                     
 \def\aas{{g^*_s}}

 \def\Tx{T_{\rm x}}       \def\Ts{T_{\rm s}}

 \def\TN{T_{_{\rm N}}}    \def\TEW{T_{_{\rm EW}}}

\newcommand{\vp}{\varphi}

\newcommand{\bref}[1]{(\ref{#1})}
\newcommand{\Lag}{{\cal L}}
\newcommand{\th}{\theta}
\newcommand{\thb}{\bar{\theta}}
\newcommand{\al}{\alpha}
\newcommand{\ald}{{\dot{\alpha}}}
\newcommand{\sm}{{\sigma^\mu}}
\newcommand{\sigb}[1]{\bar{\sigma}^{{#1}}}
\newcommand{\sgth}{\sigma^\vp}
\newcommand{\sgr}{\sigma^r}
\newcommand{\sgz}{\sigma^z}
\newcommand{\la}{\lambda}
\newcommand{\lab}{\bar{\lambda}}
\newcommand{\psb}{\bar{\psi}}
\newcommand{\phb}{\bar{\phi}}
\newcommand{\xib}{\bar{\xi}}

\newcommand{\dmu}{{\partial_\mu}}
\newcommand{\dth}{{\partial_\vp}}
\newcommand{\dr}{{\partial_r}}
\newcommand{\eth}[1]{e^{{#1}\vp}}
\newcommand{\spinU}{\mbox{\scriptsize 
		$\left(\begin{array}{c} 1 \\ 0 \end{array} \right)$}}

\newcommand{\mat}[4]{\left(\begin{array}{cc} 
			{#1} & {#2} \\ {#3} & {#4} \end{array}\right)}

\begin{document}

\title{COSMIC STRINGS IN REALISTIC PARTICLE PHYSICS THEORIES AND BARYOGENESIS}

\author{A.C. DAVIS}

\address{Department of Applied Mathematics and Theoretical Physics,}
\address{University of Cambridge,} 
\address{Cambridge, CB3 9EW, UK}


\maketitle\abstract{ 
              Grand unified theories can admit cosmic strings with
              fermion zero modes. Such zero modes result in the string
              being current-carrying and the formation of stable remnants,
              vortons. However, the string zero modes do not automatically
              survive subsequent phase transitions. In this case the vortons
              dissipate. It is possible that the dissipating cosmic vortons
              create the observed baryon asymmetry of the universe. 
              We show that fermion zero modes are an automatic consequence
              cosmic strings in supersymmetric theories. Since supersymmetry
              is not observed in nature, we consider possible supersymmetry
              breaking terms. Some of these terms result in the zero modes
              being destroyed. We calculate the baryon asymmetry generated
              by the consequent dissipating cosmic vortons. If the 
              supersymmtry breaking scale is high enough, then the dissipating
              cosmic vortons could account for the observed baryon asymmetry.
              }

\section{ Introduction}

Many particle physics theories admit cosmic strings. For most cosmological
studies the simple abelian Higgs model is used as a prototypical cosmic
string theory. However, in realistic particle physics theories the situation
is more complicated. The resulting cosmic strings can have a rich 
microstructure. Additional features can be aquired at the string core at
each subsequent symmetry breaking. This additional microstructure can, in
some cases, be used to constrain the underlying particle physics theory
to ensure consistency with standard cosmology. For example, if the theory
admits cosmic strings which aquire fermion zero modes, or bose condensates,
either at formation or due to a subsequent symmetry  then the zero modes
can be excited and will move up or down the string, depending on whether 
they are left or right movers. This will result in the string carrying 
a current \cite{witten}. An intially weak current on a string loop will be 
amplified 
as the loop contracts. The current could become sufficiently strong
to halt the contraction of the loop, preventing it from decaying. 
A stable state, or vorton \cite{vorton}, is formed. The density of vortons 
is tightly constrained by cosmological requirements. For example, if vortons 
are sufficiently stable so that they survive until the present time, then 
we require that the universe is not vorton dominated. However, if vortons
only survive a few minutes then they can still have cosmological 
implications. We then require that the universe be radiation dominated
at nucleosynthesis. 
These requirements have been used in \cite{rob&brandon} to constrain such 
models. 

Vortons are classically stable 
\cite{vortonstab}, but the quantum stability is an open question. It has 
been assumed that, if vortons decay, they do so by quantum mechanical 
tunnelling. This would result in them being very long lived. However, in the 
case of fermion superconductivity, the existence of fermion zero modes at
high energy does not guarantee that such modes survive subsequent
phase transitions. The disappearance of such zero modes could give
another channel for the resulting vortons to decay. Fermion zero modes
could also be created at subsequent phase transitions. It is thus necessary
to trace the microphysics of the cosmic string from formation through all
subsequent phase transitions in the history of the universe. 

For example, many popular particle physics theories above the electroweak scale
are based on supersymmetry. Such theories can also admit cosmic string 
solutions \cite{mark1}. Since supersymmetry is a natural symmetry between
bosons and fermions, the fermion partner of the Higgs field forming the
cosmic string is a zero mode. Thus, the particle content and interactions
dictated by supersymmetry naturally give rise to current-carrying strings.
Gauge symmetry breaking can arise either by introduction of a super-potential
or by means of a Fayet-Iliopoulos term. In both cases fermion zero modes 
arise.

However, supersymmetry is not observed in nature and must therefore be
broken. We consider general soft supersymmetry breaking terms that could
arise and consider the resulting affect of these on the fermion zero modes.
For most soft breaking terms, the zero modes are destroyed. Hence, any
vortons formed would dissipate. However, in the case of gauge symmetry
breaking via a Fayet-Iliopoulos term, the zero modes, and hence vortons,
survive supersymmetry breaking. Hence, supersymmetric theories which break
a $U(1)$ symmetry this way would result in cosmologically stable vortons,
and would therefore be ruled out. However, in the more general case, the
problem of cosmic vortons seems to solve itself. That is to say, vortons
will be formed at high energy, but will dissipate after the supersymmetry
breaking scale.

If the underlying supersymmetric theory is a grand unified one, then,
in the string core the grand unified symmetry is restored and typical
grand unified processes will be unsuppressed in the string core. Once
the vortons decay, the grand unified particles will be released.
Their out-of-equilibrium decay results in a baryon asymmetry being produced. 
Depending on the scale of supersymmetry breaking, the baryon asymmetry
produced could account for that required by nucleosynthesis.

In this talk we address this problem. We first review cosmic strings in
supersymmetric theories, displaying the string zero modes \cite{mark1}. 
We then consider
the effect of supersymmetry breaking on these zero modes, showing that
the zero modes are destroyed in the general case \cite{mark2}. 
The vortons density is estimated in these supersymmetric theories. We show
that the underlying theory can be constrained in the case where the vortons
are stable. If the vortons are unstable, we estimate 
resulting baryon asymmetry from dissipating cosmic vortons. We also 
take into account the change in entropy density from the vorton decay and 
show that, for supersymmetry breaking just before the vorton density 
dominates that of radiation, results in a baryon asymmetry, in agreement with 
observation \cite{wbp&me97}.

\section{Cosmic Strings in Supersymmetric Theories }
We consider supersymmetric versions of the spontaneously broken gauged
$U(1)$ abelian Higgs model. These models are related to or are simple
extensions of those found in reference\cite{Fayet I}. In superfield notation, 
such a theory consists of
a vector superfield $V$ and $m$ chiral superfields $\Phi_i$, ($i=1\ldots m$),
with $U(1)$ charges $q_i$. In the Wess-Zumino gauge these may be 
expressed in component notation as

\begin{equation}
V(x,\th,\thb) = -(\th\sm\thb)A_\mu(x) + i\th^2\thb\lab(x) 
		- i\thb^2\th\la(x) + \frac{1}{2}\th^2\thb^2 D(x) \ ,
\label{vectorDef}
\end{equation}

\begin{equation}
\Phi_i(x,\th,\thb) = \phi_i(y) + \sqrt{2}\th\psi_i(y) + \th^2 F_i(y) \ ,
\label{chiralDef}
\end{equation}
where $y^\mu = x^\mu + i\th\sm\thb$.
Here, $\phi_i$ are complex scalar fields and $A_\mu$ is a vector field. These
correspond to the familiar bosonic fields of the abelian Higgs model.
The fermions $\psi_{i \alpha}$, $\lab_{\alpha}$ and
$\la_{\alpha}$ are Weyl spinors and the complex bosonic fields, $F_i$, and 
real bosonic field, $D$, are auxiliary fields. Finally, $\th$ and $\thb$ are
anticommuting superspace coordinates. In the component formulation of the 
theory one eliminates $F_i$ and $D$ via their equations of motion and
performs a Grassmann integration over $\th$ and $\thb$.
Now define

\bea 
D_\al & = & \frac{\partial}{\partial\th^\al} +
i\sigma^\mu_{\al \ald} \thb^\ald \dmu \ , \nonumber  \\
\bar{D}_\ald & = & -\frac{\partial}{\partial\thb^\ald} -
i\th^\al \sigma^\mu_{\al \ald} \dmu \ , \nonumber \\
W_\al & = & -\frac{1}{4}\bar{D}^2 D_\al V \ , 
\eea
where $D_\al$ and $\bar{D}_\ald$ are the supersymmetric covariant derivatives
and $W_\al$ is the field strength chiral superfield.
The superspace Lagrangian density for the theory is then given by
\begin{equation}
{\tilde \Lag} = \frac{1}{4}
\left( W^\al W_\al|_{\th^2} + \bar{W}_\ald \bar{W}^\ald|_{\thb^2} \right) 
+ \left( \bar{\Phi}_i e^{g q_i V} \Phi_i \right)|_{\th^2\thb^2}
+ W(\Phi_i)|_{\th^2} +\bar{W}(\bar{\Phi}_i)|_{\thb^2} + \kappa D \ .
\label{susyLag}
\end{equation}
In this expression $W$ is the superpotential, a holomorphic function of the 
chiral superfields (i.e. a function of $\Phi_i$ only and not $\bar{\Phi}_i$) 
and $W|_{\th^2}$ indicates the $\th^2$ component of $W$.
The term linear in $D$ is known as the Fayet-Iliopoulos
term \cite{Fayet II}. Such a term can only be present
in a $U(1)$ theory, since it is not invariant under more general gauge
transformations. 

For a renormalizable theory, the most general superpotential is

\be
W(\Phi_i) = a_i \Phi_i + \frac{1}{2}b_{ij} \Phi_i\Phi_j 
			+ \frac{1}{3}c_{ijk} \Phi_i\Phi_j\Phi_k \ ,
\ee
with the constants $b_{ij}$, $c_{ijk}$ symmetric in  their indices.
This can be written in component form as

\be
W(\phi_i, \psi_j, F_k)|_{\th^2}
  = a_i F_i + b_{ij}\left(F_i\phi_j - \frac{1}{2}\psi_i\psi_j\right)
	+ c_{ijk} \left(F_i\phi_j\phi_k - \psi_i\psi_j\phi_k \right) 
\ee
and the Lagrangian \bref{susyLag} can then be expanded in Wess-Zumino gauge
in terms of its component fields using (\ref{chiralDef},\ref{vectorDef}).
The equations of motion for the auxiliary fields are

\be
F^\ast_i + a_i + b_{ij}\phi_j + c_{ijk}\phi_j\phi_k = 0 \ ,
\ee
\be
D + \kappa + \frac{g}{2} q_i \phb_i \phi_i = 0 \ .
\label{deqn}
\ee
Using these to eliminate $F_i$ and $D$ we obtain the Lagrangian density 
in component form as

\begin{equation}
\Lag = \Lag_B + \Lag_F + \Lag_Y - U \ ,
\label{nsusyLag}
\end{equation}
with
\begin{eqnarray}
\Lag_B &=& (D^{i\ast}_\mu \phb_i) (D^{i\mu} \phi_i)
		- \frac{1}{4} F^{\mu\nu}F_{\mu\nu} \ , \\
\Lag_F &=& -i\psi_i \sm D^{i\ast}_\mu \psb_i - i\la_i \sm \dmu \lab_i \ , \\
\Lag_Y &=& \frac{ig}{\sqrt{2}} q_i \phb_i \psi_i \la 
 	  - \left(\frac{1}{2}b_{ij} + c_{ijk}\phi_k \right) \psi_i \psi_j 
	  + (\mbox{c.c.}) \ , \\
   U   &=& |F_i|^2 + \frac{1}{2}D^2  \nonumber \\
       &=& |a_i + b_{ij}\phi_j + c_{ijk}\phi_j\phi_k|^2 
     + \frac{1}{2}\left(\kappa + \frac{g}{2} q_i \phb_i \phi_i \right)^2 \ ,
\label{Ueqn}
\end{eqnarray}
where $D^i_\mu = \dmu + \frac{1}{2}ig q_i A_\mu$ and 
$F_{\mu\nu} = \dmu A_\nu - \partial_\nu A_\mu$. 

Now consider spontaneous symmetry breaking in these theories.
Each term in the superpotential must be gauge invariant.
This implies that $a_i \neq 0$
only if $q_i =0$, $b_{ij} \neq 0$ only if $q_i + q_j =0$, and 
$c_{ijk} \neq 0$ only if $q_i + q_j + q_k=0$. 
The situation is a little more complicated than in non-SUSY theories, since
anomaly cancellation in SUSY theories implies the existence of more than one 
chiral superfield (and hence Higgs field). In order to break the gauge
symmetry, one may either
induce SSB through an appropriate choice 
of superpotential, or, in the case of the $U(1)$ gauge
group, one may rely on a non-zero Fayet-Iliopoulos term.

We shall refer to the theory with superpotential SSB (and, for simplicity, 
zero Fayet-Iliopoulos term) as theory F and
the theory with SSB due to a non-zero Fayet-Iliopoulos term as theory D. 
Since the
implementation of SSB in theory F can be repeated for more general gauge 
groups, we
expect that this theory will be more representative of general defect-forming
theories than theory D for which the mechanism of SSB is specific to the 
$U(1)$ gauge group.

\subsection{Theory F: Vanishing Fayet-Iliopoulos Term}
The simplest model with vanishing Fayet-Iliopoulos term 
($\kappa=0$) and 
spontaneously broken gauge symmetry contains three chiral superfields.
It is not possible to construct such a model with fewer superfields which 
does not
either leave the gauge symmetry unbroken or possess a gauge anomaly.
The fields are two charged fields $\Phi_\pm$, with respective $U(1)$ charges 
$q_\pm = \pm 1$, and a neutral field, $\Phi_0$. A suitable superpotential is 
then

\begin{equation}
W(\Phi_i) = \mu \Phi_0 (\Phi_+ \Phi_- - \eta^2) \label{susyW} \ ,
\end{equation}
with $\eta$ and $\mu$ real.
The potential $U$ is minimised when $F_i=0$ and $D=0$. This occurs when
$\phi_0=0$, $\phi_+ \phi_- = \eta^2$, and $|\phi_+|^2 = |\phi_-|^2$.
Thus we may write $\phi_\pm = \eta e^{\pm i\al}$, where $\alpha$ is some 
function. We shall now seek the Nielsen-Olesen\cite{NO} solution 
corresponding to an infinite straight cosmic string.  
We proceed in the same manner as for
non-supersymmetric theories. Consider only the bosonic fields (i.e. set the 
fermions to zero) and in cylindrical polar coordinates $(r,\vp, z)$ write 

\begin{eqnarray}
\phi_0 & = & 0 \ , \\
\phi_+ & = & \phi_-^\ast = \eta e^{in\vp}f(r) \ , \\
A_\mu & = & -\frac{2}{g} n \frac{a(r)}{r}\delta_\mu^\vp \ , \\
F_\pm & = & D = 0 \ , \\
F_0 & = & \mu \eta^2 (1 - f(r)^2) \ ,
\label{StringSol}
\end{eqnarray}
so that the $z$-axis is the axis of symmetry of the defect. The profile 
functions, $f(r)$ and $a(r)$, obey 

\begin{equation}
f''+\frac{f'}{r} - n^2\frac{(1-a)^2}{r^2} = \mu^2 \eta^2 (f^2 -1)f \ ,
\label{fEqn}
\end{equation}

\begin{equation}
a''-\frac{a'}{r} = -g^2 \eta^2(1-a)f^2 \ ,
\label{aEqn}
\end{equation}
with boundary conditions 

\bea
f(0)=a(0)=0 \ , \nonumber \\ 
\lim_{r\rightarrow \infty}f(r)=\lim_{r\rightarrow\infty}a(r)=1 \ .\nonumber
\eea
Note here, in passing, an interesting aspect of topological defects in
SUSY theories. The ground state of the theory is supersymmetric,
but spontaneously breaks the gauge symmetry while in the core of the defect the
gauge symmetry is restored but, since $|F_i|^2 \neq 0$ in the core, 
SUSY is spontaneously broken there. 

We have constructed a cosmic string solution in the bosonic sector of the 
theory. Now consider the fermionic sector.
With the choice of superpotential \bref{susyW} the component form of the
Yukawa couplings becomes

\begin{equation}
\Lag_Y = i\frac{g}{\sqrt{2}}
	 \left(\phb_+ \psi_+ - \phb_- \psi_-\right) \la
		 - \mu \left(\phi_0 \psi_+ \psi_- + \phi_+ \psi_0 \psi_- 
		+ \phi_- \psi_0 \psi_+ \right) + (\mbox{c.c.})
\end{equation}

As with a non-supersymmetric theory, non-trivial zero energy fermion
solutions can exist around the string. Consider the fermionic ansatz 

\be
\psi_i = \spinU \psi_i(r,\vp) \ ,
\ee
\be
\la = \spinU \la(r,\vp) \ .
\ee
If we can find solutions for the $\psi_i(r,\vp)$ and $\la(r,\vp)$ then, 
following Witten, we know that solutions of the form

\be
\Psi_i=\psi_i(r,\vp)e^{\chi(z + t)} \ , \
 \Lambda=\la(r,\vp)e^{\chi(z + t)} \ ,
\label{witteq}
\ee
with $\chi$ some function, represent left moving superconducting 
currents flowing along the string at the speed of light. Thus, the
problem of finding the zero modes is reduced to solving for the 
$\psi_i(r,\vp)$ and $\la(r,\vp)$.

The fermion equations of motion derived from \bref{nsusyLag} are four
coupled equations given by

\be
\eth{-i}\left(\dr -\frac{i}{r}\dth \right)\lab - \frac{g}{\sqrt{2}} \eta
		f \left(\eth{in}\psi_- - \eth{-in} \psi_+\right) = 0 \ ,
\label{fermeq1}
\ee
\be
\eth{-i}\left(\dr -\frac{i}{r}\dth \right)\psb_0 + i \mu \eta 
		f \left(\eth{in}\psi_- + \eth{-in} \psi_+\right) = 0 \ ,
\label{fermeq2}
\ee
\be
\eth{-i}\left(\dr -\frac{i}{r}\dth \pm n\frac{a}{r}\right)\psb_\pm +
  \eta f \eth{\mp in}\left(i\mu \psi_0 \pm \frac{g}{\sqrt{2}}\la \right) = 0 
\ .
\label{fermeq3}
\ee
The corresponding equations for the lower fermion components can be obtained 
from those for the upper components by complex conjugation, and 
putting $n \rightarrow -n$. The superconducting current corresponding to this 
solution (like \bref{witteq}, but with $\chi(t-z)$) is right moving.

We may enumerate the zero modes using an index theorem \cite{index}, as 
discussed further in \cite{stephen}. This gives $2n$ independent zero
modes, where $n$ is the winding number of the string. However, in
supersymmetric theories we can calculate them explicitly using SUSY 
transformations. This relates the fermionic components of the superfields to 
the bosonic ones and we may use this to obtain the fermion solutions in terms 
of the background string fields. A SUSY transformation is 
implemented by the operator $G=e^{\xi Q + \xib \bar{Q}}$, 
where $\xi_{\alpha}$ are Grassmann parameters and $Q_{\alpha}$ are the 
generators of the SUSY algebra which we may represent by

\begin{eqnarray}
Q_\al & = & \frac{\partial}{\partial\th^\al} 
- i\sigma^\mu_{\al \ald} \thb^\ald \dmu \ , \\
\bar{Q}^\ald & = & \frac{\partial}{\partial\thb_\ald}
		- i\bar{\sigma}^{\mu \ald \al} \th_\al \dmu \ .
\label{SusyTransform}
\end{eqnarray}

In general such a transformation will induce a change of gauge. It is then 
necessary to perform an additional gauge transformation to return to the
Wess-Zumino gauge in order to easily interpret the solutions. For an abelian 
theory, supersymmetric gauge transformations are of the form

\begin{eqnarray}
\Phi_i & \rightarrow & e^{-i\Lambda q_i}\Phi_i \ , \\
\bar{\Phi}_i & \rightarrow & e^{i\bar{\Lambda} q_i}\bar{\Phi}_i \ , \\
V & \rightarrow & V + \frac{i}{g}\left(\Lambda - \bar{\Lambda}\right) \ ,
\end{eqnarray}
where $\Lambda$ is some chiral superfield. 

Consider performing an infinitesimal SUSY transformation on \bref{StringSol},
using $\dmu A^\mu = 0$. The appropriate $\Lambda$ to return to Wess-Zumino 
gauge is

\begin{equation}
\Lambda = ig\xib\sigb{\mu}\th A_\mu (y) 
\end{equation}

The component fields then transform in the following way

\begin{eqnarray} 
\phi_\pm(y) & \rightarrow & \phi_\pm(y)  
+ 2i\th\sm\xib D_\mu \phi_\pm(y) \ , \\ 
\th^2 F_0(y) & \rightarrow & \th^2 F_0(y) + 
2\th\xi F_0(y) \ , \\
-\th\sm\thb A_\mu(x) & \rightarrow &  
	-\th\sm\thb A_\mu(x) \nonumber \\
&& {}+ i\th^2 \thb\frac{1}{2} \sigb{\mu}\sigma^\nu \xib F_{\mu \nu}(x) 
- i\thb^2 \th\frac{1}{2} \sm\sigb{\nu} \xi F_{\mu \nu}(x)
\ .
\end{eqnarray}
Writing everything in terms of the background string fields,
only the fermion fields are affected to first order by the transformation.
These are given by

\begin{eqnarray}
\la_\al &\rightarrow& \frac{2na'}{gr}i(\sgz)^\beta_\al \xi_\beta \ , \\
(\psi_\pm)_\al &\rightarrow& \sqrt{2} \left(if'\sgr \mp
\frac{n}{r}(1-a)f \sgth\right)_{\al \ald} \xib^\ald \eta \eth{\pm in} \ , \\
(\psi_0)_\al &\rightarrow& \sqrt{2}\mu\eta^2(1-f^2)\xi_\al \ ,
\end{eqnarray}
where we have defined

\begin{eqnarray}
\sgth & = & \mat{0}{-i\eth{-i}}{i\eth{i}}{0} \ , \\
\sgr & = & \mat{0}{\eth{-i}}{\eth{i}}{0} \ .
\end{eqnarray}

Let us choose $\xi_{\alpha}$ so that only one component is nonzero.
Taking $\xi_2 = 0$ and $\xi_1 = -i\delta/(\sqrt{2}\eta)$, where $\delta$
is a complex constant, the fermions become

\begin{eqnarray}
\la_1 & = & \delta\frac{n\sqrt{2}}{g\eta}\frac{a'}{r} \ , \\ 
(\psi_+)_1 & = & \delta^\ast \left[f'+\frac{n}{r}(1-a)f\right]\eth{i(n-1)} 
\ , \\
(\psi_0)_1 & = & -i\delta\mu\eta(1-f^2) \ , \\
(\psi_-)_1 & = & \delta^\ast \left[f'-\frac{n}{r}(1-a)f\right]\eth{-i(n+1)} \ .
\end{eqnarray}
It is these fermion solutions which are responsible for the string
superconductivity.
Similar expressions can be found when $\xi_1 = 0$. It is clear from
these results that the string is not invariant under supersymmetry,
and therefore breaks it. However, since $f'(r), a'(r), 1-a(r)$ and $1-f^2(r)$ 
are all approximately zero outside of the string core, the SUSY breaking and
the zero modes are confined to the string. We note that this method gives us 
two zero mode solutions. Thus, for a winding number one string, we obtain the
full spectrum, whereas for strings of higher winding number, only a partial
spectrum is obtained.

The results presented here can be extended to non-abelian gauge theories.
This is done in \cite{mark2}. The results are very similar to those presented
here, so we leave the interested reader to consult the original paper.

\subsection{Theory D: Nonvanishing Fayet-Iliopoulos Term}

Now consider theory D in which there is just one primary charged chiral
superfield involved in the symmetry breaking and a non-zero Fayet-Iliopoulos
term. In order to avoid gauge anomalies, the model must contain other 
charged superfields. These are
coupled to the primary superfield through terms in the superpotential such that
the expectation values of the secondary chiral superfields are dynamically 
zero. The secondary superfields have no effect on SSB and are invariant under
SUSY transformations. Therefore, for the rest 
of this section we shall concentrate on the primary chiral 
superfield which mediates the gauge symmetry breaking. 

Choosing $\kappa = -\frac{1}{2}g\eta^2$, the theory is spontaneously
broken and there exists a string solution obtained from the ansatz

\begin{eqnarray} 
\phi & = & \eta e^{in\vp}f(r) \ , \\
A_\mu & = & -\frac{2}{g} n \frac{a(r)}{r}\delta_\mu^\vp \ , \\
D & = & \frac{1}{2}g\eta^2 (1-f^2) \ , \\
F & = & 0 \ .
\end{eqnarray}
The profile functions $f(r)$ and $a(r)$ then obey the first order equations

\begin{equation}
f' = n\frac{(1-a)}{r}f
\label{fEqnII}
\end{equation}

\begin{equation}
n\frac{a'}{r} = \frac{1}{4}g^2 \eta^2 (1-f^2)
\label{aEqnII}
\end{equation}

Now consider the fermionic sector of the theory and perform a SUSY
transformation, again using $\Lambda$ as the gauge function to return to
Wess-Zumino gauge. To first order this gives

\begin{eqnarray}
\la_\al &\rightarrow&  \frac{1}{2}g\eta^2(1-f^2)i(I+\sgz)_\al^\beta\xi_\beta\\
\psi_\al &\rightarrow& 
\sqrt{2}\frac{n}{r}(1-a)f(i\sgr-\sgth)_{\al \ald} \xib^\ald \eta \eth{in}  
\end{eqnarray}
If $\xi_1=0$ both these expressions are
zero. The same is true of all higher order terms, and so the string is
invariant under the corresponding transformation. For other $\xi$,
taking $\xi_1= -i\delta/\eta$ gives
\bea 
\la_1  &=& \delta g \eta(1-f^2) \\
\psi_1 &=& 2\sqrt{2} \delta^\ast \frac{n}{r}(1-a)f \eth{i(n-1)} 
\eea
Thus supersymmetry is only half broken inside the string. This is in
contrast to theory F which fully breaks supersymmetry in the string
core. The theories also differ in that theory D's zero
modes will only travel in one direction, while the zero modes of theory F 
(which has twice as many) travel in both directions. In both theories the
zero modes and SUSY breaking are confined to the string core.

Thus, a necessary feature of cosmic strings in SUSY theories is that 
supersymmetry is broken in the string core and the resulting strings have
fermion zero modes. As a consequence, cosmic strings arising in SUSY theories
are automatically current-carrying. In general, cosmic strings arise as
infinite strings or as closed loops, The usual non current-carrying string
loops decay via gravitational radiation. However, in current-carrying 
strings loops do not necessarily suffer the same fate. The loops could be
stabilised by the angular momentum of the current carriers, forming a stable,
vorton, configuration. Vortons are classically stable objects 
\cite{vortonstab}, though their quantum mechanical stability of an open
question. The presence of vortons puts severe constraints on the underlying 
theory since the density of vortons could overclose the universe if vortons
are stable enough to survive to the present time. 
If they only live for a few minutes then the vorton density could affect
nucleosynthesis. This is discussed in detail in \cite{rob&brandon}. However, 
in some theories the vorton problem solves itself.

\section{Soft Susy Breaking}
Supersymmetry is not observed in nature. Hence, it must be broken. 
Supersymmetry breaking is achieved by adding soft SUSY breaking terms which do
not induce quadratic divergences.  

In a general model, one may obtain soft SUSY breaking terms by the 
following prescription.

\begin{enumerate}
\item Add arbitrary mass terms for all scalar particles to the scalar 
potential.

\item Add all trilinear scalar terms in the superpotential, plus
their hermitian conjugates, to the scalar potential with arbitrary coupling.

\item Add mass terms for the gauginos to the Lagrangian density.
\end{enumerate}

Since the techniques we have used are strictly valid 
only when SUSY is exact, it is necessary to investigate the effect of these 
soft terms on the fermionic zero modes we have identified.

As we have already commented, the existence of the zero modes can be seen as 
a consequence of an index theorem~\cite{index}. The index is
insensitive to the size and exact form of the Yukawa couplings, as
long as they are regular for small $r$, and tend to a constant at
large $r$. In fact, the existence of zero modes relies only on
the existence of the appropriate Yukawa couplings and that they have the 
correct $\vp$-dependence. Thus there can only be a change in the number of zero
modes if the soft breaking terms induce specific new Yukawa couplings
in the theory and it is this that we must check for. Further, it was
conjectured in~\cite{index} that the destruction of a zero mode occurs only 
when the relevant fermion mixes with another massless fermion. 

We have examined each of our theories with respect to this criterion and list 
the results below.

\subsection{Theory-F}

As discussed previously, the superpotential for this theory is,
\be
W = \mu \Phi_0 (\Phi_+ \Phi_- - \eta^2) \ .
\ee
The trilinear and mass terms that arise from soft SUSY breaking
are
\be
m_0^2 |\phi_0|^2 + m_-^2 |\phi_-|^2 + m_+^2 |\phi_+|^2 
 + \mu M \phi_0\phi_+\phi_-
\ee
The derivative of the scalar potential with respect to $\phi_0^\ast$ becomes
\be
\phi_0 (\mu^2|\phi_+|^2 + \mu^2|\phi_-|^2 + m_0^2) + \mu M(\phi_+ \phi_-)^\ast
\ee
This will be zero at a minimum, and so $\phi_0 \ne 0$ only if $M \ne 0$.

New Higgs mass terms will alter the values of $\phi_+$ and $\phi_-$ slightly, 
but will not produce any new Yukawa terms. Thus these soft
SUSY-breaking terms have no effect on the existence of the zero modes.

However, the presence of the trilinear term gives $\phi_0$ a non-zero 
expectation value, which gives a Yukawa term coupling the $\psi_+$ and $\psi_-$
fields. This destroys all the zero modes in the theory since the left
and right moving zero modes mix. 

For completeness note that a gaugino mass term also mixes the left and right 
zero modes, aiding in their destruction.

\subsection{Theory-D}

The $U(1)$ theory with gauge symmetry broken via a Fayet-Iliopoulos term and 
no superpotential is simpler to analyse. New Higgs mass terms 
have no effect, as in the above  case, and there are no trilinear
terms. Further, although the gaugino mass  terms affect the form of
the zero mode solutions, they do not affect their  existence, and so,
in theory-$D$, the zero modes remain even after SUSY breaking. For 
this class of theories, the strings remain current-carrying and, hence,
have a potential vorton problem. This could lead to the theories being in
comflict with cosmology.

\section{Current-Carrying Strings and Vortons}

For the theories considered in the previous sections, the strings become
current-carrying due to fermion zero modes as a consequence of supersymmetry. 
These zero modes are present in the string core at formation. If we call
the temperature of the phase transition forming the strings $\Tx$, we can
estimate the vorton density. The more general case to consider would be
when the string becomes current-carrying at a subsequent phase transition,
but this is beyond the scope of this paper and we refer the reader to 
\cite{mark2}.

The string loop is characterised by two currents, the topologically conserved
phase current and the dynamically conserved particle number current. Thus
the string carries two conserved quantum numbers; $N$ is topologically
conserved integral of the phase current and $Z$ is the particle number.
A non conducting Kibble type string loop must
ultimately decay by radiative and frictional drag processes until it
disappears completely. However, a conducting
string loop may be saved from
disappearance by reaching a state in which the energy attains a minimum for
given non zero values of $N$ and $Z$. 

It should be emphasised that the existence of such vorton states does not
require that the carrier field be gauge coupled. If there is indeed a 
non-zero charge coupling then the loop will have a corresponding total
electric charge, $Q$, such that the particle number is $Z=Q/e$. However,
the important point is that, even in the uncoupled case where $Q$ vanishes
the particle number $Z$ is perfectly well defined.

The physical properties of a vorton state is determined by the quantum 
numbers, $N$ and $Z$. However, these are not arbitrary. For example,
to avoid decaying completely like a non conducting loop, a conducting loop 
must have a non zero value for at least one of the numbers $N$ and $Z$. In 
fact, one would expect that both these numbers should be reasonably large 
compared with unity to diminish the likelihood of quantum decay by barrier 
tunneling. There is a further restriction on the 
values of their ratio $Z/N$ in order to avoid spontaneous 
particle emission as a result of current saturation. In this contribution
we are going to consider the special case where $\vert Z\vert \approx N $.
These are the so called chiral vortons.

For chiral vortons we have,

\be
\Ev\simeq \lv \mx^{\,2}\ \ . 
\ee

In order to evaluate this quantity all that remains is to work out $\lv$. 
Assuming that vortons are approximately circular, with radius
given by $\Rv=\lv/2\pi$ and angular momentum
quantum number $J$ given \cite{C ring} by $J=NZ$. Thus, eliminating $J$, 
one obtains 

\be
\lv\simeq(2\pi)^{1/2}\vert NZ\vert^{1/2}\mx^{-1} \ . 
\ee

Thus we obtain an estimate of the vorton mass energy as

\be
\Ev\simeq(2\pi)^{1/2}\vert NZ\vert^{1/2}\mx\approx N\mx\ ,
\label {energy}
\ee

where we are assuming the classical description of the string dynamics.
This is valid only if the length $\lv$ is large
compared with the relevant quantum wavelengths. This will only be satisfied 
if the product of the quantum numbers $N$ and $Z$ is
sufficiently large. A loop that does not satisfy this requirement will never
stabilise as a vorton. 

We can now calculate the vorton abundance. Assuming that the string becomes
current carrying at a scale $\Tx$ by fermion zero modes then
one expects that thermal fluctuations will give rise to a non zero value for
the topological current, $\vert j\vert^2$. Hence, a random walk process
will result in a spectrum of finite values for the corresponding
string loop quantum numbers $N$ and $Z$. Therefore, loops for which these
numbers satisfy the minimum length condition will become vortons.
Such loops will ultimately be able to survive as 
vortons if the induced current, and consequently $N$ and $Z$ are 
sufficiently large, such that

\be
\vert NZ \vert^{1/2} \gg {1}.
\ee

Any loop that fails to satisfy this
condition is doomed to lose all its energy and disappear.

The total number density of small loops with length and radial
extension of the order of $L_{\rm min}$, the minimum length for vortons,
will be not much less than the number density
of all closed loops and hence

\be
n\approx \nu\  L_{\rm min}^{-3}
\ee

where $\nu$ is a time-dependent parameter. The typical length scale of
string loops at the transition temperature, 
$L_{\rm min}(\Tx)$, is considerably greater than relevant thermal 
correlation
length, ${\Tx}^{-1}$, that characterises the local current
fluctuations. 
It is because of this that string loop evolution is
modified after current carrier condensation. Indeed, since 
$L_{\rm min}(\Tx)\gg {\Tx}^{-1}$ and loops present at the time
of the condensation satisfy $L\geq L_{\rm min}(\Tx)$, the 
random walk
effect can build up reasonably large, and typically comparable initial values
of the quantum numbers $\vert Z\vert$ and  $N$. The expected root mean square 
values produced in this way from carrier field fluctuations
of wavelength $\lambda$ can be estimated as

\be
\vert Z\vert \approx N \approx \sqrt{L\over\lambda} ,
\ee

where $\lambda\approx\Tx^{-1}$. Thus, one obtains 

\be
\vert Z\vert \approx N \approx \sqrt{L_{\rm min}(\Tx)\Tx} \gg 1 , 
\ee

For current condensation during the friction dominated regime 
this requirement is always satisfied. 

Therefore, the vorton mass density is

\be
\rhv\approx {N \mx \nv} .
\ee

In the friction dominated regime the string is interacting with the 
surrounding plasma. We can estimate $L_{\rm min}$ in this regime as the
typical length scale below which the microstructure is smoothed \cite
{rob&brandon}. This then gives the quantum number, $N$

\be
N\approx\left({\mP\over\beta\Tx}\right)^{1/4} ,
\label{old 21}
\ee

where $\beta$ is a drag coefficient for the friction dominated era that is
of order unity. We then obtain the number density of mature vortons

\be
\nv\approx\nua \left({\beta\Tx\over \mP}\right)^{3/2} T^3,
\label{plus 26}
\ee
This gives the resulting mass density of the relic vorton population to be

\be
\rhv\approx \nua \left({\beta\Tx\over\mP}\right)^{5/4}\Tx T^3 \ . 
\label{plus 31}
\ee

\subsection{The Nucleosynthesis Constraint.}

One of the most robust predictions of the standard cosmological model is the
abundances of the light elements that were fabricated during primordial
nucleosynthesis at a temperature $\TN\approx 10^{-4} \GeV$.

In order to preserve this well established picture, it is necessary that the
energy density in vortons at that time, $\rhv(\TN)$ should have been small
compared with the background energy density in radiation, 
$\rhN\approx\aa\TN^4$, where $\aa$ is the effective number of degrees of 
freedom. Assuming that carrier condensation occurs during the
friction damping regime and that $\aa$ has dropped to a value of order unity
by the time of nucleosynthesis, this gives

\be
\nua\aas^{-1}\beta^{5/4}\mP^{-5/4}\Tx^{\,9/4}\ll\TN\ .
\label{old 24}
\ee 

The case for which strings become current-carrying at formation
has been studied previously and yields
rather strong restrictions for very long lived vortons \cite{C}. If
it is only assumed that the vortons survive for a few minutes, which is all
that is needed to reach the nucleosynthesis epoch we obtain a much weaker
restriction.

\be
\left(\frac{\nua}{\aas}\right)^{4/9}\Tx \ll 
\left(\frac{\mP}{\beta}\right)^{5/9}\TN^{4/9}\ .
\label{old 25}
\ee
Taking $\aas \approx 10^2$ yields the inequality 

\be
\Tx\leq (\nua)^{-4/9}\beta^{-5/9} \times 10^9\ \GeV \ .
\label{plus 34}
\ee
This is the condition that must be satisfied by the formation temperature 
of {\it cosmic strings that become superconducting immediately}, subject to 
the rather
conservative assumption that the resulting vortons last for at least a few
minutes. If we assume that the net efficiency
factor $(\nua)^{-4/9}$ and drag factor $\beta^{-5/9}$ are of order 
unity this condition rules out the formation of such strings during any 
conceivable GUT transition, but is consistent with their formation at 
temperatures close to that of the electroweak symmetry breaking transition.

\subsection{The Dark Matter Constraint.}

Let us now consider the rather stronger constraints that can be 
obtained if at
least a substantial fraction of the vortons are sufficiently stable to last
until the present epoch.
It is generally accepted that the virial equilibrium of galaxies and
particularly of clusters of galaxies requires the existence of a cosmological
distribution of ``dark" matter. This matter must have a density 
considerably in excess of the baryonic matter
density, $\rhb\approx 10^{-31}$ gm/cm$^3$. On the other hand, on the same
basis, it is also generally accepted that to be consistent with the
formation of structures such as galaxies it is necessary
that the total amount of this ``dark" matter should not greatly exceed the
critical closure density, namely

\be
\rhc\approx 10^{-29} {\rm gm \ cm^{-3}} \ .
\label{add 15}
\ee
As a function of temperature, the critical density scales like the entropy 
density so that it is given by

\be
\rhc(T)\approx \aa\mc T^3\ ,
\label{plus 35}
\ee
where $\mc$ is a constant mass factor. For comparison with the density of 
vortons that were formed at a scale $\Tx$ we can estimate this to be

\be
\aas \mc\approx 10^{-26}\mP\approx 10^2\,\hbox{eV}\ .
\label{plus 37}
\ee
The general dark matter constraint is

\be
\Ov \equiv {\rhv\over\rhc}\leq 1\ .
\label{plus 38}
\ee

In the case of vortons formed as a result of condensation during the friction
damping regime
the relevant estimate for the vortonic dark matter fraction is obtainable
from (\ref{plus 31}) as

\be
\Ov\approx\beta^{5/4}\left({\nua\mP\over\aas\mc}\right)
\left({\Tx\over\mP}\right)^{9/4} .
\label{old 27}
\ee

The formula (\ref{old 27}) is applicable to the case considered in earlier
work \cite{C}, in which it was supposed that vortons sufficiently stable 
to last until the present epoch, with the strings becoming current-carrying
at formation, as in the case of supersymmetric theories. In this case
one obtains,

\be
\beta^{5/9}{\Tx\over\mP} \left({\nua\mP\over\aas\mc}\right)^{4/9}
\leq 1\ .
\label{plus 39}
\ee
Substituting the estimates above we obtain

\be
\Tx \leq (\nua)^{-4/9}\beta^{-5/9} \times 10^7\, \GeV\  .
\label{plus 40}
\ee

This result is based on the assumptions that the vortons in question are 
stable enough
to survive until the present day. Thus, this constraint is naturally more 
severe
than its analogue in the previous section. It is to be
remarked that  vortons produced in a phase transition occurring at or near the
limit that has just been derived would give a significant contribution to the
elusive dark matter in the universe. However, if they were produced at the
electroweak scale, i.e. with $\Tx\approx\Ts \approx\TEW$, where $\TEW\approx
10^2\, \GeV$, then they would constitute such a small dark matter fraction,
$\Ov\approx 10^{-9}$, that they would be very difficult to detect.

These constraints are very general for long lived vortons. However, if the
microphysics of the underlying theory is such that the fermion zero modes
are destroyed by subsequent phase transitions, then an entirely different
situation pertains. For example, in our F-type SUSY theory, the zero modes
didnot survive supersymmetry breaking. In this case, the current, and hence
the resulting vortons, would dissipate. We turn to this case in the next
section. If the zero modes do survive SUSY breaking, as in the case of
our D-type theory, then the theory faces a vorton problem. It seems 
possible that such theories are in conflict with observation.

\section{Dissipating Cosmic Vortons}

In general, SUSY breaking occurs at a fairly low energy. In which case a
sizeable random current will have built up in the string loops, resulting 
from string self-intersections and intercommuting. When the string 
self-intersects or intercommutes there is a finite probability that
the fermi levels will be excited. This produces a distortion in the
fermi levels, resulting in a current flow. 
As a consequence, vortons will form prior to SUSY breaking. 

For strings that are formed at a temperature $T_{\rm x}$  and become 
superconducting at formation, the vorton number density is 
\be
\nv=\nu_* \bigl({\beta T_{\rm x} \over m_{\rm P}}\bigr)^{3\over 2} T^3,
\ee
while the vorton mass density is 
\be
\rhv=\nu_* \bigl({\beta T_{\rm x} \over m_{\rm P}}\bigr)^{5\over 4}
T_{\rm x} T^3,
\ee
where $\nu_*$ and $\beta$ are factors of order unity.

In the F-type theory, the zero modes do not survive SUSY breaking. As a
conequence, the current decays, angular momentum is lost and the vorton
shrinks and eventually decays. As the vortons decay, grand unified particles
are released from the string core. Since these GUT particles are also 
unstable, they also decay, but in a baryon number violating manner. As they
decay, they create a net baryon asymmetry. 

Given the number density of vortons at the SUSY breaking transition we can 
estimate the baryon asymmetry produced by vorton decay using,
\be
{n_b \over s}={\nv \over s}\epsilon K,
\ee
where $s$ is the entropy density, $\epsilon$ is the baryon asymmetry produced  by a 
GUT particle and $K$ is the number of GUT particles per vorton. 
We need to consider two cases: firstly the vortons may decay before they 
dominate the
energy density of the universe and we do not need to know the time scale for 
vorton decay since $\nv/s$ is an invariant quantity. Alternatively, if the 
vorton energy density does dominate the energy density of the 
universe we must modify the temperature evolution of the universe to allow 
for entropy generation.

Assuming that the universe is radiation dominated until after the electroweak 
phase transition, the temperature of the 
universe is simply that of the standard hot big bang. We can estimate the 
entropy density following vorton decay using the standard result,
\be
s={2\pi^2 \over 45} g^* T^3,
\ee
where $g^*$ is the effective number of degrees of freedom at the electroweak 
scale ($\simeq 100$). The vorton to entropy ratio is then

\be
{\nv \over s} \simeq \bigl({T_{\rm x}\over m_{\rm P}}\bigr)^{3\over 2}{45\over 2\pi^2 g^*}
\sim 5\times 10^{-6},
\ee
for $T_{\rm x}\sim 10^{16}$GeV.

The number of GUT particles per vorton is obtained from (\ref{energy})
\be
K=\bigl({\beta T_{\rm x} \over m_{\rm P}}\bigr)^{-{1\over 4}}
\sim 10,
\ee
and we have 
\be
{n_b \over s}\sim 10^{-5}\epsilon.
\ee

Alternatively, the vorton energy density may come to dominate and we must allow for a non-standard temperature evolution.
The temperature of vorton-radiation equality, $T_{\rm veq}$, is given by
\be
T_{\rm veq}={\nu_* \over g^*} \bigl({\beta T_{\rm x} \over m_{\rm P}}\bigr)^{5\over 4}T_{\rm x}.
\ee
If we assume that the vortons decay at some temperature $T_d$ and reheat the 
universe to a temperature $T_{\rm rh}$, we
 have
\be
\hat g^*  T_{\rm rh}^4=\rhv(T=T_{\rm d})=
\nu_* \bigl({\beta T_{\rm x} \over m_{\rm P}}\bigr)^{5\over 4}
T_{\rm x} T_{\rm d}^3,
\ee
where $\hat g^*$ is the number of degrees of freedom for this lower 
temperature. This reheating and  entropy generation leads to an extra baryon 
dilution factor. In this case the baryon asymmetry produced by the decaying 
vortons is given by
\be
{n_b\over s}={\nv\over s} K\epsilon[{g^*\over \hat g^*}{ T_{\rm eq}\over T_{\rm d}}] ^{-{3\over 4}},
\ee
where the entropy, $s$,  is that of the standard big bang model. The universe 
now evolves as in the standard big bang model and 
$n_b/s$ remains invariant. Using the above results the asymmetry becomes,

\be
{n_b\over s}=\epsilon(\nu^* {\hat g^{*3}\over g^{*'4}})^{1\over4}\beta^{5\over16}
\bigl({ T_d^{12}\over m_{\rm P}^5 T_{\rm x}^7}\bigr)^{1\over16}.
\ee

This form is valid if the vortons dominate the energy density of the universe 
before they decay, if this is not the case the dilution factor is absent 
and we have
\be
{n_b\over s}\simeq{\epsilon\over g^{*'}}\bigl({ T_{\rm x}\over m_{\rm P}}\bigr)^{5\over4} ,
\ee
as above.

The maximal asymmetry produced is if the vortons decay just before they 
dominate the energy density. This requires $T_d \ge 10^6 Gev$ for grand 
unified strings. In this case, since $\epsilon$ is of order $0.01$ in
many GUT theories, the mechanism can easily produce 
\be 
{n_b\over s}\simeq{10^{-10}},
\ee
as required by nuclesynthesis.

\section{Discussion}
In this contribution we have considered the microphysics of cosmic strings
arising in physical particle physics theories. In the first part we 
concentrated on cosmic strings in supersymmetric theories, uncovering many
novel features, including the possibility of them carrying persistent currents.
We then considered the fate of current-carrying string loops, showing that
they form stable vortons. We were able to use this to constrain the underlying
theory. We then considered the possibility, suggested in the first part,
that the vortons could dissipate and, in doing so, create the observed
baryon asymmetry.
 
In particular we investigated the structure of cosmic string solutions to
supersymmetric abelian Higgs models. For completeness we have analysed two 
models, differing
by their method of spontaneous symmetry breaking. However, we expect theory F 
to be more representative of general
defect forming theories, since the SSB employed there is not specific to 
abelian gauge groups. 

We have shown that although SUSY remains unbroken outside the
string, it is broken in the string core (in contrast to the gauge
symmetry which is restored there). In theory F supersymmetry is broken 
completely in the string core by
a nonzero $F$-term, while in theory D supersymmetry is partially
broken by a nonzero $D$-term. We have demonstrated that, due to the
particle content and couplings dictated by SUSY, the cosmic
string solutions to both theories are superconducting in the Witten sense. 
We believe this to be quite a powerful result, that all supersymmetric
abelian cosmic strings are superconducting due to fermion zero modes. An 
immediate and important application of the results of the present paper is 
that SUSY GUTs 
which break to the standard model and yield abelian cosmic strings (such as 
some breaking schemes of $SO(10)$) must face strong constraints from 
cosmology\cite{rob&brandon}. 

While we have performed this analysis for an abelian 
string, the techniques to be quite general and the results for non-abelian
theories are very similar. 

We have also analysed the effect of soft SUSY breaking on the existence 
of fermionic zero modes. The Higgs mass terms did not affect the
existence of the zero modes. In the theories with $F$-term symmetry
breaking, gaugino mass terms destroyed all zero modes which involved
gauginos and trilinear terms created extra Yukawa couplings which
destroyed all the zero modes present. 
In the theory with $D$-term symmetry breaking, the zero modes
were unaffected by the SUSY breaking terms. If the remaining zero modes survive
subsequent  phase transitions, then stable vortons could result. Such
vortons would dominate the energy density of the universe, rendering
the underlying GUT cosmologically problematic.

Therefore, although SUSY breaking may alleviate the cosmological disasters 
faced by superconducting cosmic strings~\cite{rob&brandon}, there are classes 
of string solution for which zero modes remain even after SUSY breaking. It
remains to analyse all the phase transitions undergone by specific SUSY
GUT models to see whether or not fermion zero modes survive down to the
present time. If the zero modes do not survive SUSY breaking, the
universe could experience a period of vorton domination beforehand,
and then reheat and evolve as normal afterwards.

We then went on to calculate the remnant vorton density, assuming that
the strings become current-carrying at formation, as is the case for
the supersymmetric theories under consideration. We used this density
to constrain the underlying theory for the case of a persistent current.
Two separate cases were considered. If the vortons survive for only
a few minutes, we demanded that the universe be radiation dominated 
throughout nucleosynthesis to constrain the scale of symmetry breaking
to be less than $10^{9}$ GeV. However, if the vortons survive to the
present time, then we can demand that they do not overclose the universe.
In this case we obtained a much stronger constraint that the scale of 
symmetry breaking must be less than $10^{7}$ GeV. This suggests that
GUT theories based on D-type supersymmetric theories, which would 
automatically predict the existence of cosmic strings with the properties we 
have uncovered, are in conflict with observation.  

On the constructive side, we have
shown that it is possible for various conceivable symmetry breaking
schemes to give rise to a remnant vorton density sufficient to make up a
significant portion of the dark matter in the universe.

We have also shown that vortons can decay after a subsequent phase transition 
and these dissipating vortons can create a baryon asymmetry. For example,
the zero modes in the F-type theories do not automatically survive SUSY
breaking. In this case, the decaying vortons could account for the observed
baryon asymmetry, depending on the scale of supersymmetry breaking.
If the SUSY breaking scale were just above the electroweak scale, then
the resulting asymmetry may well not be enough. This is due to the fact that 
vortons dominate the energy density of the universe long before they decay. 
Their decay results in a reheating of the universe and an increase in the 
entropy  density. This reheating is unlikely to have any effect on the 
standard cosmology following the phase transition. If however the
scale of SUSY breaking were such that the vortons didnot dominate the energy
density of the universe, then their decay could explain the observed baryon
asymmetry of the universe.

\section*{Acknowledgments}
I wish to thank my collaborators:- Robert Brandenberger, Brandon Carter,
Stephen Davis, Warren Perkins and Mark Trodden for fruitful and enjoyable
collaborations. This work was supported in part by PPARC.
Finally, I wish to thank Edgard Gunzig for inviting me to such a stimulating 
meeting and Mady Smet for allowing us to use her wonderful venue of Peyresq.
 

\end{document}